# Time-lapse, three-dimensional in situ imaging of ice crystal growth in a colloidal silica suspension

Sylvain Deville [a,1], Jérôme Adrien [b], Eric Maire [b], Mario Scheel [c], Marco Di Michiel [c]

[a] Laboratoire de Synthèse et Fonctionnalisation des Céramiques, UMR3080 CNRS/Saint-Gobain, Cavaillon, France

[b] Université de Lyon, INSA-Lyon, MATEIS CNRS UMR5510, Villeurbanne, France

[c] ESRF, Experimental Division, Grenoble, France

**Abstract**

The freezing of colloidal suspensions is encountered in many natural and engineering processes such as the freezing of soils, food engineering, and cryobiology. It can also be used as a bio-inspired, versatile, and environmentally-friendly processing route for porous materials and composites. Yet, it is still a puzzling phenomenon with many unexplained features, due to the complexity of the system and the space and time scales at which the process should be investigated. We demonstrate here the interest of fast X-ray computed tomography to provide time-lapse, three-dimensional, in situ imaging of ice crystal growth in a colloidal silica suspension. The experimental measurements show that the local increase of colloid concentration does not affect the growth kinetics of the crystals, until the colloidal particles become closely packed. For particles much smaller than ice crystals, the concentrated colloidal suspension is equivalent to a simple liquid phase with higher viscosity and a freezing point determined by the concentration of colloidal particles.

**Keywords**: solidification, solid/liquid interface, synchrotron radiation computed tomography, freeze-casting, pattern formation

---

[1] Corresponding author. Email : sylvain.deville@saint-gobain.com

# 1. Introduction

The solidification or freezing of colloidal suspension or colloids is commonly encountered in a variety of natural processes such as the freezing of soils in northern regions and the growth of sea ice [1], or everyday life and engineering situations such as food engineering [2], materials science [3,4], cryobiology [5], filtration or water purification and the removal of pollutants from waste [6]. It also has implications in various materials engineering processes such as the processing of particle-reinforced alloys and composites [7] or the processing of porous materials using the freezing of colloidal suspension, a process referred to as ice templating or freeze casting [3,8,9]. The phenomenon by itself is surprisingly simple to describe: a solidification interface, usually the water/ice interface, is propagating through a colloidal suspension of particles, cells or micro-organisms. This simplicity is nevertheless misleading and we are still far from complete understanding and control of the phenomenon. A very large number of disparate parameters should be accounted for when trying to understand and model the freezing of colloids [10]. The characteristics of the colloidal suspension are often critical to the behaviour of the system during freezing and to a large extent have never been analysed nor understood.

A solid experimental foundation on which further theoretical developments can be built and validated is required to better understand colloid freezing. Precise and quantitative observations are required for understanding the process. Current approaches, both direct –optical microscopy [11], X-ray radiography [12] – or indirect –X-ray scattering [13] – only provide partial, averaged or indirect observations. Information on the ice crystal morphology can also be obtained by looking at their replica, obtained for instance through the ice-templating approach. Samples have been freeze-dried to remove the ice and consolidated by sintering. Such observations are therefore indirect and cannot provide kinetic information about the phenomena. Ideally, we need in situ, real time, three-dimensional observations of both the crystal's growth and colloid's movement in the suspensions, with individual colloid tracking and concentration measurement to investigate the colloid redistribution during freezing. The growth velocity of individual crystals, perpendicular to the main solidification direction, is a good example of a measurement that cannot be achieved so far (Figure 1). The concentration of particles between adjacent crystals increases with the thickness of the crystals, which should in turn affect the growth velocity of the crystals.

The corresponding space (submicron) and velocity (several to hundreds of $\mu m.s^{-1}$) scales severely restrict the choice of experimental techniques. The crystal growth velocity can be controlled by the cooling gradient, so that the phenomenon can be slowed down to be investigated. The velocity range of interest (1-50 $\mu m.s^{-1}$), corresponding to the usual occurrences of colloids freezing, nevertheless requires high resolution, fast imaging capabilities. X-ray computed tomography [14] can fulfill most of these requirements: transparent materials are not required because the imaging is based on absorption contrast and a high spatial resolution (1-2 µm) can be obtained. It was successfully applied in solidification studies to investigate the solidification of alloys [15–17] or the frozen structure of colloids [12]. Nevertheless, a complete investigation of the kinetics-structure relationships in freezing colloids was not accessible until now, the technique being limited by the time required to perform a tomography scan.

Here, we demonstrate that fast X-ray computed tomography allows time-lapse, three-dimensional in situ imaging of ice crystal growth in a colloidal silica suspension. We take advantage of recent upgrades and developments on the beamline ID15A at the ESRF [18], which, due to the high flux of high energy X-rays available, allow us to perform a complete tomography acquisition within a second, instead of several minutes in our previous attempts [14].

## 2. Methods

### 2.1 Materials and freezing setup

A commercial colloidal silica aqueous suspension of 100nm particles was used (DP5540, Nyacol, Ashland, MA, USA). The suspension concentration was 30wt. %. The freezing setup used has been described in previous papers [12]. It comprised a tube placed over a copper cold finger, cooled by circulation of liquid nitrogen. A heater and temperature controllers provide good control of the cooling conditions and thus of the freezing kinetics. The mould is insulated from the atmosphere to avoid condensation of moisture at its outer surface. Ice crystal growth occurs along the direction of the temperature gradient, referred to in the paper as the vertical or z-direction. The horizontal plane corresponds to the xy plane, perpendicular to the vertical direction. A cooling rate of 2°C.min$^{-1}$ was used.

### 2.2 Fast X-ray computed tomography

Synchrotron absorption micro-tomography measurements were performed at the ID15A beam line of ESRF using a pink 31keV X-ray beam with a bandwidth of 6keV. Fast tomography scans were performed using a high resolution imaging detector. The detector consists of a 15µm thick LuAG:Ce$^{2+}$ scintillator that converts the X-ray absorption signal into a visible light image, which is then magnified by 10× optics and recorded by a high speed 12 bit CCD camera. The image pixel size is 1.786 × 1.786 µm$^2$. Thanks to the high flux of high energy radiation provided by the ID15 undulator source, 3D tomographic scans of the ice crystal growth and particle redistribution could be acquired in situ with the pink beam in 1.5s, limiting irradiation of the sample. 500 projections were obtained for each scan, with an exposure time of 3.5ms per projection. Tomographic reconstructions were performed using the supercomputer at ESRF.

The acquisition frequency –one tomography scan every 12.5s– is limited by the characteristics of the cooling stage. The circulating nitrogen used to cool the stage is brought in by tubes. The presence of these tubes currently prevents a continuous rotation of the stage. The movement of the stage comprised several steps in our experiments:

- an initial acceleration to reach the rotation velocity required for the scan. This occurs over around 30°.
- a steady state, where the rotation velocity is constant, during which the tomography acquisition is actually performed, over 180°.
- a progressive decrease of the rotation velocity until the stage stops. This occurs over around 30°.
- a return to the initial position of the stage for the next acquisition.

In the future, having a continuous rotation would therefore allow us to remove all the intermediate stages. This could increase the acquisition frequency by a factor of approximately five to ten. Unfortunately this was not available for the present experiment but it could be designed for future developments.

## 3. Results

### 3.1 Time-lapse observations

A first run of experiments was conducted by scanning the samples at the highest possible frequency. As explained above, a tomography scan was obtained every 12.5s. The corresponding 3D reconstructions of the ice crystals are shown in figure 2. Because the diameter of the mould (3 mm) is larger than the dimensions of the region scanned (around 900µm), the reconstructions show the central part of the sample only (local tomography mode). Possible effects due to interactions between the suspension and mould walls are therefore not observed in these images. The time-lapse reconstruction provides a good overview of the progressive growth of the ice crystals along the solidification direction (z).

Heterogeneities in the horizontal growth of the ice crystals (horizontal stripes) can clearly be observed, in the form of regularly spaced dendrites (ridges) (figure 3). Dendrites form in specimens not subjected to the beam irradiation, as described in almost every paper on freeze casting. The fundamental difference here is the orientation of the dendrites. In all papers published so far, dendrites are oriented along the direction of the temperature gradient, see for instance references [4,19–24]. In the current observations, the dendrites are oriented perpendicular to the direction of the temperature gradient. The occurrence of these dendrites is synchronised with that of the tomography scans and is therefore very probably related to the presence of the X-ray beam.

The role of sample rotation during the scan was assessed, for it could induce some shear in the suspension, in particular close to the crystals tips. Acquisition were performed with and without rotation (beam open), and conversely with the beam closed and with rotation. In the latter case, a snapshot is taken after the rotation is completed. This lead us to the conclusion that the rotation of the stage has no influence on the characteristics of the ice crystals morphology, as we observed that horizontal dendrites appear even if the acquisition is performed without rotation. These horizontal dendrites therefore arise from irradiation by the beam.

Another run of experiments was performed by making a scan every 120s, so as to let the crystals recover their equilibrium regime. These time-lapse reconstructions are shown in figure 4. The region of the crystals affected by the beam during the previous scan is now located 400-500µm below the crystals tip (figure 5). Although less information can be obtained about the dynamics of the process from such experiments, the crystal morphologies observed are much less affected by the beam and thus better reflect normal conditions during freezing. Dendrites –vertical ridges observed along the vertical growth direction– can be observed on the side of the crystals (figure 6). They are observed for all the crystals, growing along the solidification direction. These morphologies are in good agreement with the morphologies observed in ice-templated materials, where the particles replicate the ice crystal morphology [4].

The observations thus provide an accurate quantitative description of the crystal morphology and the growth of the crystals relative to each other. During the freezing of colloidal suspensions, ice crystals usually exhibit a lamellar growth morphology with a dendritic surface on one side. The surface of the opposite side is usually smooth and free of dendrites. This peculiar morphology is linked to the growth anisotropy of ice and the low tilt of the ice crystals [25]. Adjacent ice crystals usually share the same orientation, with the flat side facing the same direction. In figure 4, ice crystals with different orientations can be observed first in the upper left part of figure 4a. These crystals then progressively cross the observation window, due to their slight tilt. Despite presenting an opposing orientation, their growth is unaffected by the presence of the surrounding crystals.

## 3.2 Quantitative analysis

Due to the growth anisotropy and the mismatch between the preferred growth direction and the temperature gradient direction [25], ice crystals in colloids usually grow with a small tilt, with respect to the solidification direction. The reconstructions shown in figure 4 can be used to measure the tilt of the crystals in situ. The results, plotted in figure 7, reveal that the tilt of the crystals is related to the growth kinetics. When performing a scan every 120s, the stage is moved between the observations, so that the actual growth velocity cannot be precisely measured. Nevertheless, the position of the beam-induced heterogeneities, discussed previously and illustrated in figure 5, can be used to monitor indirectly the growth of the ice crystals between consecutive scans (figure 7b). Tilting of the ice crystals becomes greater when the growth velocity decreases. A threshold, below which ice crystals tilt noticeably, exists in the 3-3.5µm.s$^{-1}$ range.

A quantitative analysis of the crystals' morphology can also be performed. The thickness of the ice crystals can be measured quantitatively, by an erosion/dilatation method performed on the cross section perpendicular to the main growth direction (z) [26]. Its evolution with time is plotted in figure 8a. With a constant cooling rate of 2°C.min$^{-1}$ applied at the bottom of the sample, the crystals growth speed progressively decreases. This results in a progressive increase of the ice crystal thickness. Under the same conditions, characterisation of the dendrites shows (figure 7b) that their periodicity is related to the thickness of the crystals. Dendrites spacing becomes larger as crystals grow thicker.

Using the fastest acquisition frequency, additional information can be obtained, although the influence of the beam must be taken into account. In particular, the top position of the ice crystals with time can be precisely monitored so that the position and vertical growth velocity along the temperature gradient direction –vertical direction– can be tracked. The crystal tip position vs. time is plotted in figure 9a. An average vertical crystal growth velocity of 1.58µm.s$^{-1}$ ($R^2$=0.996) can be estimated for the upper part of the sample from the slope of the position vs. time data. Measuring the instantaneous vertical growth velocity by the local derivative of the position (figure 9b) can provide indications about the growth regime. The instantaneous growth velocity data reveal oscillatory instabilities (figure 9c), a situation that has been predicted theoretically [27]. The oscillations are particularly visible in the upper part of the sample. These oscillations are not on the same scale as the acquisitions or rotations.

Using the data from figure 9a and the measurements of the ice crystal thicknesses, the thickness of the ice crystals can be tracked over time (figure 10a). Although the very first moments are not accessible, the thickness linearly increases with time, and then progressively slows down to reach a maximum value. The slope in the initial linear part of the thickness vs. time data can be used to locally estimate–at a defined z position– the horizontal growth velocity. The results are shown in figure 10b. The crystal's lateral growth velocity shows an oscillatory behaviour and a progressive increase with time. The oscillations can be related to the lateral heterogeneities induced by the beam, illustrated in figure 10c. The progressive increase of the lateral growth velocity is correlated to a progressive decrease of the growth velocity along the temperature gradient direction (vertical). This eventually leads to thicker ice crystals, in agreement with the thickness data shown in figure 8.

Combining the local thickness measurements (averaged for all the crystals in the xy plane) at each z position and for each scan with the crystal tip position data as shown in figure 9a, we can obtain a master curve of the ice crystal thickness evolution versus time. The results were obtained by linearly extrapolating the position between two consecutive data points (figure 9b), so that the precise time at which the ice crystals tip reach a z position can be accurately estimated. The average crystal thickness vs. time depends on the growth kinetics along the temperature gradient direction (vertical). Two examples are discussed here. In figure 11a, the plot corresponds to the lower part of the sample, where the vertical growth velocity (3-4µm.$s^{-1}$) is larger than in the upper part (around 1.6µm.$s^{-1}$) (figure 9c). The maximum thickness of the crystals is thus lower, around 30µm. The thickness data show oscillations, which are artefacts apparently related to the lateral growth of heterogeneities, such as those shown in figure 10c. In figure 11b, similar data are shown for the upper part of the sample, where the vertical growth velocity is lower and the maximum thickness of the crystals reaches 40µm. The crystals apparently experience less heterogeneities as evidenced by less significant oscillations in the thickness data.

An intermediate plateau of thickness can be observed in figure 11a (arrow), where the thickness of the ice crystals is constant for some time, around 10µm, and then eventually increases (a schematic representation is plotted in figure 11c). This plateau can be related to the observation of the crystal's tips, shown in figure 11d. The tip of the crystals shown here, and representative of the rest of the crystals, reveal an uppermost section with a small radius of curvature, around 11µm, and then increases to a larger radius of curvature. This morphology is typical of those obtained with the beam for an acquisition frequency of one tomography scan every 12.5s. When performing the experiments with a frequency of one scan every 120s, the part of the tip with this small radius of curvature disappears. It is thus a direct consequence of the beam on the behaviour of the system, and is further discussed later in the paper. When the vertical growth velocity decreases, the tip with the small radius of curvature is still present (figure 11e), but less important, and with an even lower radius of curvature, around 7µm. The intermediate plateau in the thickness vs. time data (figure 11b) is also less visible.

## 4. Discussion

These unique imaging capabilities offered by fast tomography brought substantial information on the freezing process of colloids and the associated mechanisms. What is new and unique here can be gathered in three main categories:

- Qualitative observations: the temporal morphology evolution of the ice crystals growing into a colloidal suspension have never been observed before. Our observations reveal, for instance, the specific morphology of the tip of the crystal, which is strongly asymmetric and does not change with time. This is very different from the usual dendritic morphologies observed in solidification, where the tip is highly symmetrical.

- Quantitative observations: growth kinetics is measured both along the temperature gradient and perpendicular to the temperature gradient, which have never been obtained thus far. We can now compare the growth kinetics in both directions (parallel and perpendicular to the temperature gradient) and conclude that they are actually quite similar. This is an important result since the lateral growth kinetics dictates the conditions for particle redistribution between the crystals. These values of growth kinetics can now be used as input in modeling approaches to obtain more realistic predictions (see for instance [28]). Such qualitative and quantitative observations can be compared to the output of phase field or level set models, although such models are yet to be developed for the solidification of colloidal suspensions. Finally, we also investigate the development of secondary dendrites and the tilt of the crystals, both qualitatively and quantitatively, and link these to the growth kinetics.

- X-rays computed tomography is becoming a very popular and powerful technique for solidification studies [15,17,29,30]. These results highlight some limitations of the technique, in particular regarding the influence and absorption of the beam. We demonstrate here that aqueous systems are highly sensitive to small variations of temperature (a few degrees), which can drastically affect the solidification behaviour and resulting morphologies.

These three aspects are discussed in the following paragraphs.

## 4.1 Possibilities and system adaption using fast X-ray tomography

To obtain a good computation of the structure from the set of radiographs, the evolution of the structure during the scan must be, at most, on the order of the spatial resolution. With a spatial resolution of 1.78µm, this implies that the growth velocity must be approximately lower than 2-5µm.s$^{-1}$. The measured vertical growth velocity in these experiments was set to be very close to these conditions (figure 8c). These values, however, are slightly lower than the growth velocities used in the ice-templating process, which are typically in the range 10-50µm.s$^{-1}$. The ice crystals morphology observed here is similar to what is observed in the standard conditions and this suggests that the growth behaviour is essentially the same.

Among solidification studies, the water-to-ice phase transformation is likely one of the most tricky to investigate. The system is highly sensitive to perturbations, such as additional energy brought in by the irradiation of the beam. Small fluctuations of the temperature or the temperature gradient can drastically affect the experimental observations. The fast X-ray computed tomography approach presented here offers a unique opportunity to investigate the phenomenon. Additional modification can nevertheless be considered to further improve the investigations. The difference in absorption is sufficient to discriminate between the ice and the regions of concentrated particles. Ideally, since the absorption of the signal is proportional to the concentration of particles, we should also be able to quantify the concentration of particles. Although the intensity of the beam is high, the acquisition time of a single radiograph is short (3.5ms), and thus the signal intensity is not strong enough to be sensitive to absorption variations induced by concentration fluctuations. Several improvements can be considered to enable such measurements, such as further progress in the sensitivity of the detectors, increase of the number or projections or alternatively decrease the magnification. Such results would be highly beneficial to a greater understanding of the process and the underlying mechanisms. Local tomography and phase contrast should also be minimized as these are known to affect the relation between density and attenuation.

The results also revealed the influence of the beam on the behaviour of the system. The intensity of the beam being very high (approximately $4\times10^{13}$ photons/s.mm$^2$), we can estimate that samples under irradiation by the beam get a heating rate of about 2°C.s$^{-1}$ (assuming that the whole sample absorbs like water). The sample is inside the beam for 2-3s for each tomography scan. The thermal gradient and undercooling of the tip, which controls the radius of curvature, are therefore affected by the beam. The occurrence of these heterogeneities are initially synchronised with the presence of the X-ray beam (e.g. seven heterogeneities are observed with the seventh scan), and then the synchronisation is lost. In addition, the results also reveal oscillations of the growth velocity, which indicate an unstable growth regime. We can therefore speculate that, in addition to heating the sample, the energy brought to the system by the beam sends the crystals into an unstable regime, characterized by the growth of horizontal dendrites. In this case, the frequency of the lateral dendrites is dictated by the growth regime and not the acquisition frequency. Such behaviour can only be resolved by using a lower beam intensity, which implies a detector with a better sensitivity. In the meantime, we have shown that decreasing the acquisition frequency can partially solve this issue. If the acquisition occurs at a low frequency (such as one scan every 120s), the crystals recover very fast –within a few seconds– so that artefact-free observation can be obtained. The observations are nevertheless at a much lower frequency and yield less information about the dynamic behaviour of the system. Intermediate growth stages cannot be observed, the lateral growth of the ice crystals cannot be tracked over time. The morphology of the crystals in these conditions cannot be related to their growth kinetics.

The current procedure for measuring the lateral growth velocity is also dependent on the acquisition and experimental conditions. In these experiments, an initial linear regime of around 60s was observed, corresponding to four or five data points. This is enough to obtain a reliable measurement of the slope, which gives the growth velocity. Faster growth velocity will provide less data points and therefore less reliable measurements, unless the acquisition frequency can be increased.

## 4.2 Qualitative and quantitative observations

The lamellar morphology and directionality of the structure of the current observations is less regular and homogeneous than what is seen and presented in other papers. Several parameters contribute to this behaviour:

- What is usually observed is the replica of the ice crystals, i.e. the porous structure. Getting a feel of the ice crystals morphology require imagining the replica of the porous structure. Here we have a direct observation of the crystals. The appreciation can thus be slightly different from what is usually seen.

- Usually, the morphology is observed after freeze-drying and sintering, which both affects the morphology. Here, the observations are performed in situ, so that the morphologies are not affected by the subsequent steps.

- As described extensively in the paper, the absorption of the beam by the sample during freezing affects the development of secondary dendrite. These secondary dendrites are not observed in other papers, the morphologies are thus quite different, as described above.

- The growth kinetics are a little bit slower that those used when shaping materials by ice templating. At lower growth velocity, crystals and secondary dendrites can grow to larger dimensions (see fig. 8b), and the distribution of the size of the crystals becomes larger.

- The imposed cooling rate is lower than the usual processing conditions for ice-templated materials. This results in a smaller temperature gradient. The driving force imposing a single growth direction is thus weaker.

- Finally, we exert no control of the nucleation conditions. The crystals generally grow with a short-range order (adjacent crystals are parallel to each other), and with a small tilt. When a crystal is presenting a different tilt, as observed in figure 4, it will thus cross the observation window with time. Crystals are thus apparently growing in different directions. This becomes more apparent when the tilt of the crystals increases.

The observations unaffected by the beam, where a scan was performed every 120s, revealed the relationships between the tilt of the ice crystals and the kinetics conditions. For the fast growth velocity (faster than $3µm.s^{-1}$), the tilt of the ice crystals is very small, around 1°. It is thus difficult to observe and measure it on ice-templated samples, but it is more important and thus measurable in our observations. When the vertical growth velocity resulting from the imposed temperature gradient diminishes, the tilt of the crystals progressively increases. Looking at the data in figure 7, there seems to be a velocity threshold below which the tilt is constant, and above which the tilt progressively increases. Even if the tilt is small, it impacts the morphology of the ice crystals. The ice crystals usually do not exhibit a flat surface and dendrites grow laterally. When the crystals are tilted, the dendrites grow only on one side, with respect to the temperature gradient. It is energetically unfavourable to grow on the other side.

This behaviour is related to the crystallographic orientation of the crystals. Previous in situ X-Ray diffraction results [31] indicated that under the usual freezing conditions, the c-axis of the hexagonal ice crystals is almost aligned with the direction of the temperature gradient. The actual growth direction results from the competition between the direction imposed by the temperature gradient (z) and the direction imposed by the ice crystal growth anisotropy (figure 12). The results obtained here indicate that when the temperature gradient is large, and hence the growth velocity is high ($>3.5µm.s^{-1}$), the temperature gradient has a predominant influence for selecting the growth direction. Below the observed velocity threshold, the direction imposed by the crystal growth anisotropy becomes more important as the temperature gradient decreases, resulting in a larger tilt of the crystals.

Measurements of the lateral growth velocity (figure 9 and 10) reveal that the initial stages of growth are characterized by a constant growth velocity. The concentration of colloids in the intercrystal space increases as crystals grow laterally. The particle concentration and freezing point are correlated, although not linearly [32]. It eventually reaches a threshold where the concentration of particles is so high that the freezing point decreases locally very fast, while the high concentration of particles slows down further growth of the crystals. In addition, the lateral growth velocity apparently increases as the vertical growth velocity –along the temperature gradient– decreases (figure 10b). Such behaviour can now be incorporated into more realistic models of colloid freezing [28,33,34]. The freezing behaviour of concentrated colloids results from the interplay of many parameters. In particular, the interaction between a concentrated colloidal suspension and the growing crystals has long been questioned. From the results presented here, it is clear that a local increase of colloid concentration does not affect the growth kinetics of the crystals, until the colloidal particles are closely packed. The colloidal particle size probably plays a critical role. In this study, the colloidal particle size is around 100nm, and the typical thickness of ice crystals is around 30µm, that is, 300 times larger than particles. As long as the colloid size is much smaller –one or two orders of magnitude– than the ice crystal size, individual colloids are too small to exert a noticeable influence on the ice crystals. The concentrated colloidal suspension is equivalent to a simple liquid phase with higher viscosity, and a freezing point determined by the concentration of colloidal particles. Replicating similar experiments with both larger (1µm and smaller (10nm) particles is required to determine the size range within which this approximation is valid.

In its current configuration, the method is certainly of interest for several aspects of the mechanisms of the freezing of colloids investigated. In particular, the situation of low growth velocity (1µm.s$^{-1}$ or less), which can be accessed here, typically corresponds to the ice lens growth situation, observed frequently in geophysics [35,36]. Time-lapse three-dimensional observation of ice lens growth would be highly beneficial to the current understanding of the process. One of the unresolved issues is the location of ice lens nucleation, so ice lenses first nucleate in frozen or unfrozen concentrated colloidal suspension, or in large pores, or even branch off pre-existing lenses? Beyond this particular situation, the data obtained here can now be used as input for the different models of the freezing of colloids currently developed, such as linear stability analysis [37], discrete element modelling (molecular dynamics [28]) or phase field models [38].

## 5. Conclusions

The recent progress in X-ray computed tomography at the ESRF provides access to rapid time-lapse, three dimensional, in situ imaging, which we used to investigate the growth of ice crystal in a colloidal suspension. We illustrate here with a colloidal silica suspension both qualitative and quantitative morphological and kinetic information that can be obtained with such experiments. The experimental results show that the local colloid concentration increase does not affect the growth kinetics of the crystals, until colloidal particles become closely packed. For particles much smaller than ice crystals, the concentrated colloidal suspension is thus equivalent to a simple liquid phase with higher viscosity, and a freezing point determined by the concentration of colloidal particles. There are currently no alternatives to obtain such information, which prove to be critical for our understanding of this complex process.

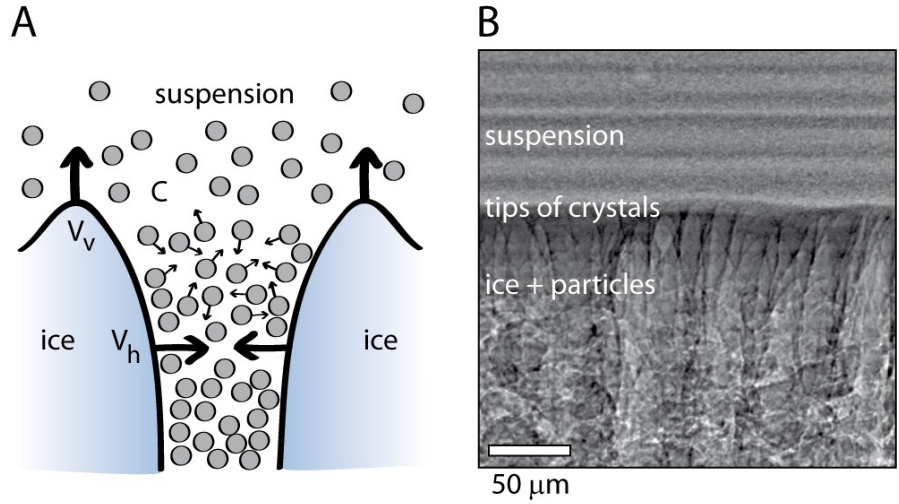

Figure 1. Schematic representation of crystal growth and particle redistribution (A). The vertical ($V_v$) and horizontal ($V_h$) growth velocity are represented by the arrows. The particles are mostly repelled by the interface moving in the horizontal direction. A typical radiograph obtained by conventional X-rays radiography [39] is shown in (B). The tips of crystals are all found at the same height, measuring $V_v$ from such radiographs is therefore straightforward. $V_h$ cannot be obtained using current approaches. Scale bar: 50 µm.

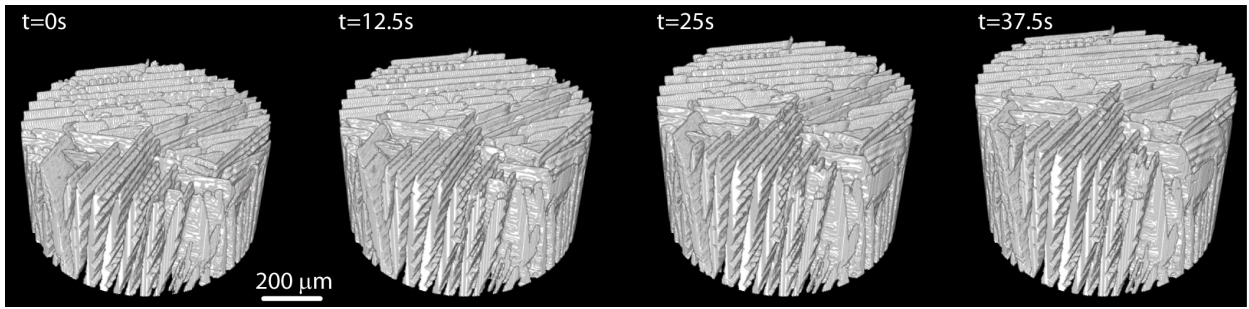

Figure 2. Time lapse, 3D reconstruction of the ice crystals. The diameter of the reconstruction region is 860µm.

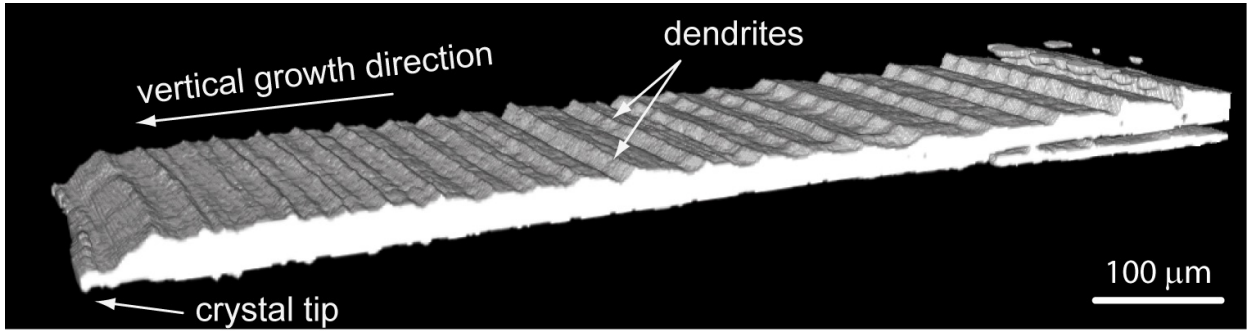

Figure 3: Close up view of an isolated crystal, from tomography scans taken every 12.5s, showing the dendritic artefacts (ridges) growing perpendicular to the vertical growth direction. Reconstructed volume is $50\times220\times900\mu m^3$.

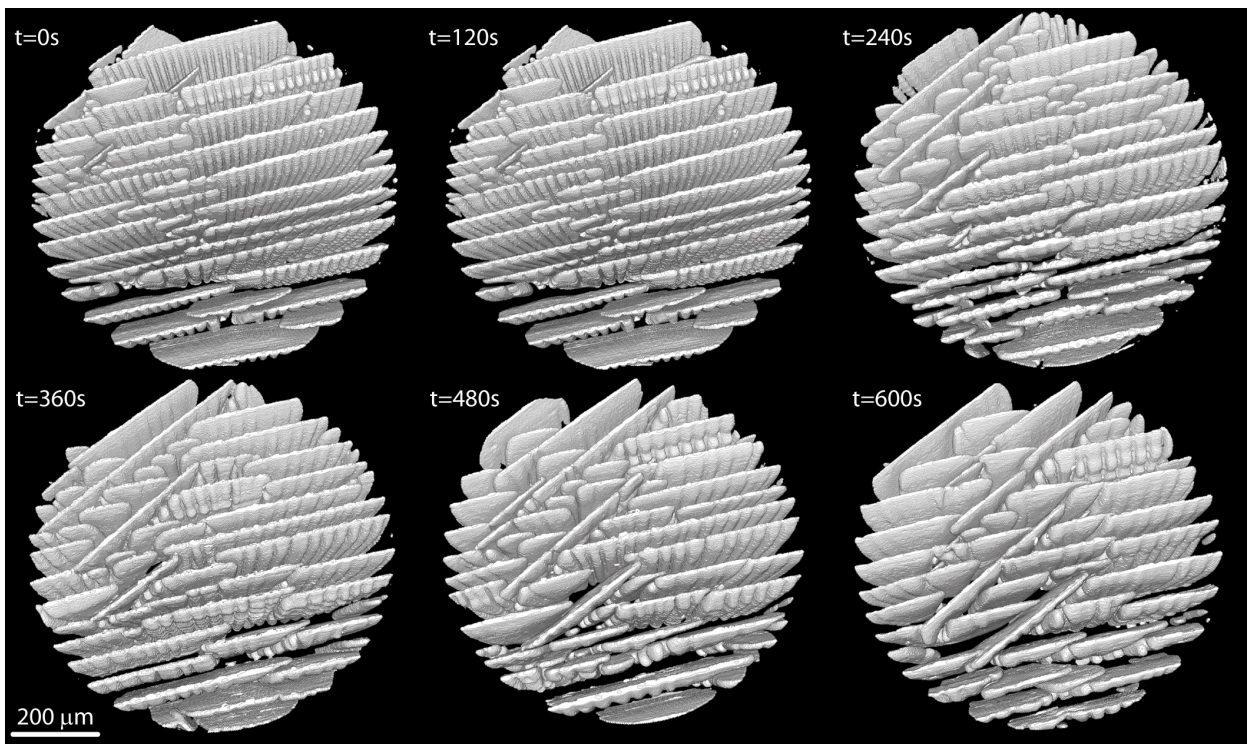

Figure 4. Time lapse, 3D reconstruction of the ice crystals, seen from above. The diameter of the reconstruction region is 860µm.

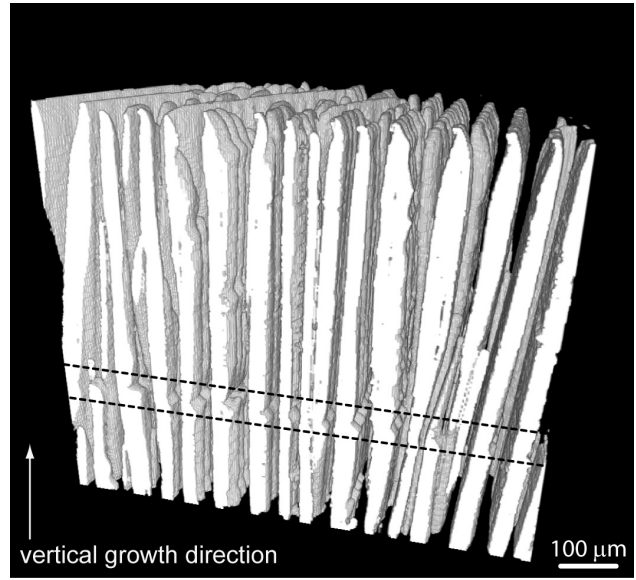

Figure 5: Close up view of a 3D reconstruction of the ice crystals, seen from aside, from tomographs taken every 120s. The dashed lines indicate the approximate location of the region affected by the beam during the previous acquisition, located in this case about 500µm below the crystal's tips. The local change in thickness in this region can be observed. The diameter of the reconstruction region is 860µm.

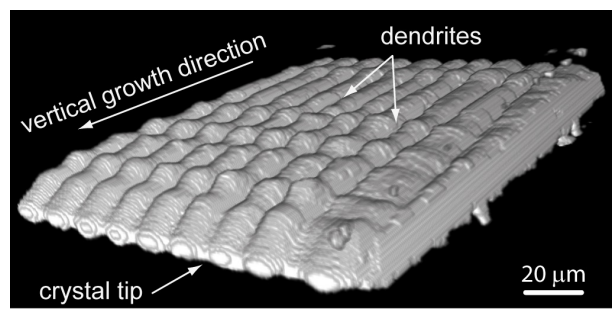

Figure 6: Close up view of an isolated crystal, from tomography scan taken every 120s, showing the dendrites (ridges) growing along the vertical growth direction. Reconstructed volume is 60×330×430µm³. The region affected by the beam during the previous acquisition, located far below the tip, is not visible here.

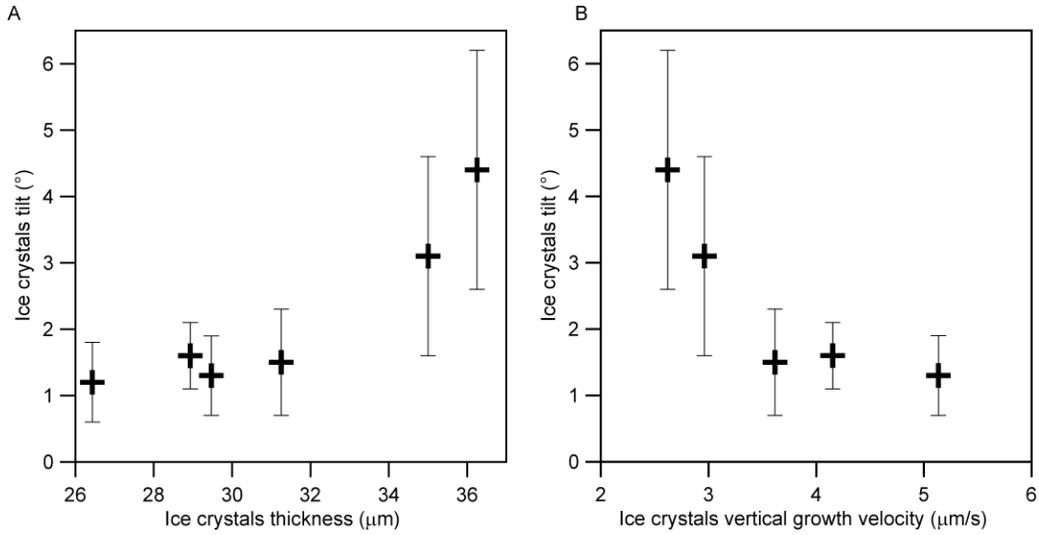

Figure 7. Ice crystals tilt vs. crystals thickness (A) and vs. ice crystals vertical growth velocity, for an acquisition frequency of one tomography scan per 120s.

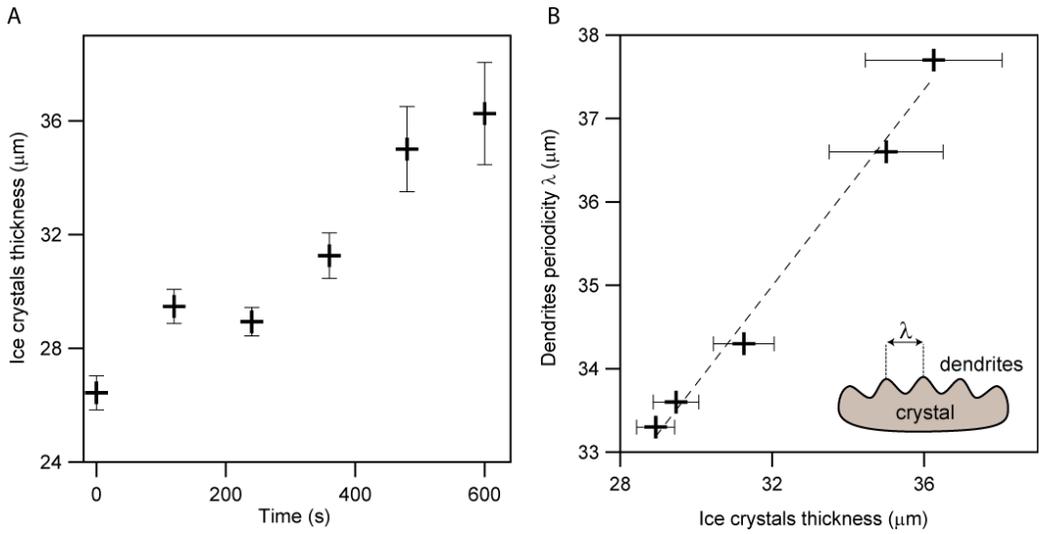

Figure 8. Time evolution of morphological characteristics of the crystals: ice crystals thickness vs. time (A) and dendrites periodicity vs. ice crystals thickness (B).

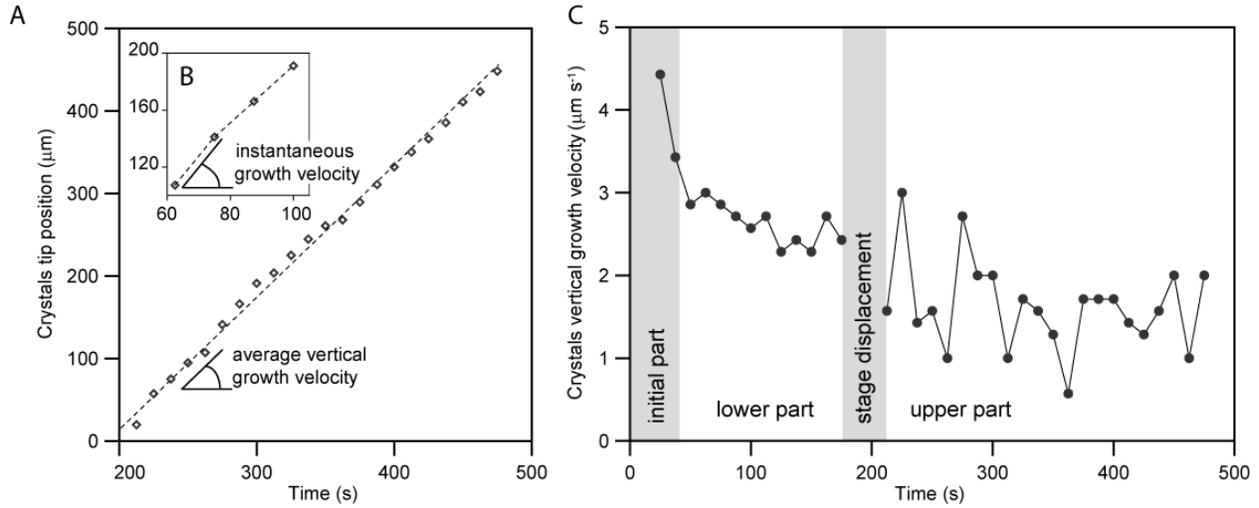

Figure 9. Crystal tip position vs. time in the upper part of the sample (A), partial view (B) of the position vs. time data showing the procedure for instantaneous growth velocity measurements and corresponding instantaneous growth velocity along the vertical direction in the sample (C). The average crystal vertical growth velocity from the position vs. time data in the upper part of the sample is 1.58µm.s$^{-1}$. The crystals experience velocity oscillations, particularly visible in the upper part of the sample i.e. when the freezing speed becomes smaller. The vertical growth velocity progressively slows down due to the reduced temperature gradient.

a

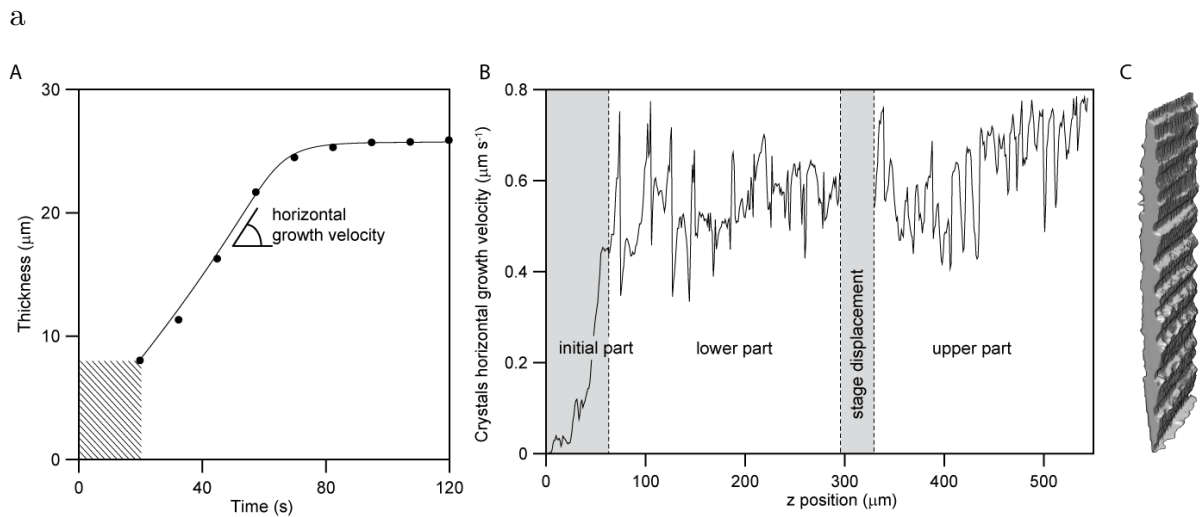

Figure 10. Thickness, horizontal growth velocity and corresponding morphologies of ice crystals. The horizontal growth velocity can be measured at each z position, from the thickness vs. time data (A). The horizontal growth velocity can thus be tracked for the entire sample (B). Velocity fluctuations reflect the growth of dendrites (C), induced by the beam. The isolated crystal shown in C is 500µm high. Due to a high vertical growth

velocity (>3µm.s$^{-1}$), there are not enough data points in the initial part of the sample to determine a proper value of the horizontal growth velocity. Values in the initial part must thus be discarded.

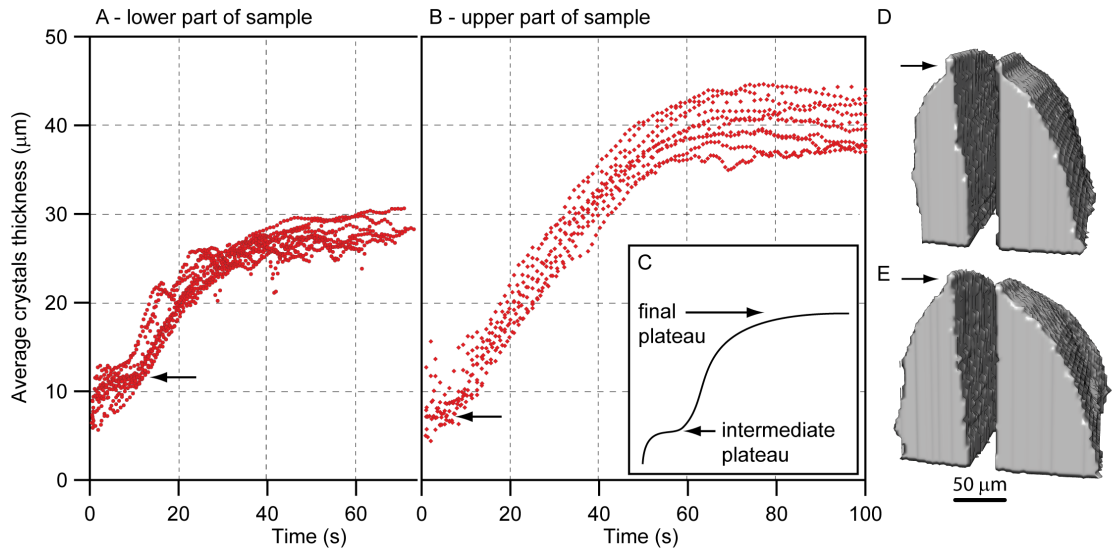

Figure 11. Master curve of ice crystals thickness evolution with time, in the lower and upper part of the sample. An intermediate plateau can be observed (arrow), corresponding to the growth of the tip with the low radius of curvature, shown in D and E (arrow). A schematic of the profile is plotted in the inset C. The intermediate plateau is more visible at high crystal growth velocity (lower part); the height of this part of the crystal is greater (comparison can be made with panels D and E). The corresponding crystals tip morphology is shown on the right. The crystals shown are 200µm high.

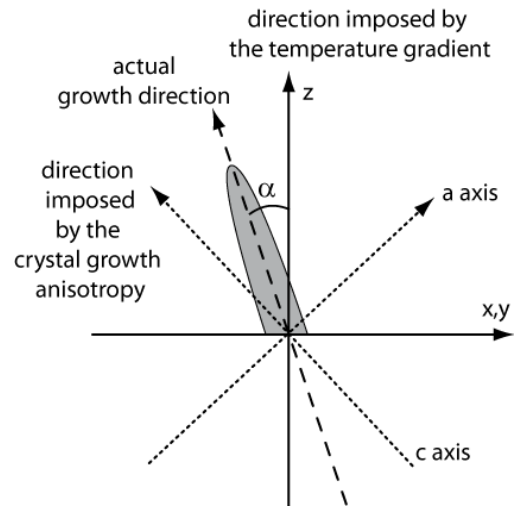

Figure 12. Actual growth direction, temperature gradient direction and crystallographic orientation of the crystals. The measured tilt is α.

**Supplementary Online Materials**

Movie S1: Time lapse, 3D reconstruction of the ice crystals growth. The diameter of the reconstruction region is 860µm. A picture is taken every 12.5s.